\documentclass[preprint, aps, prx, amsmath, superscriptaddress]{revtex4-1}

\usepackage[hidelinks,
    pdfauthor={C. Arnold, O. Vendrell, R. Santra},
    pdftitle={Electronic decoherence following photoionization: full quantum-dynamical treatment of the influence of nuclear motion}
]{hyperref}

\usepackage{units}
\usepackage{braket}
\usepackage{dsfont}

\usepackage{graphicx}
\graphicspath{ {/} {./graphics/} }

\newcommand{\chem}[1]{\ensuremath{\mathrm{#1}}}
\newcommand{\vect}[1]{\ensuremath{\mathbf{#1}}}

\begin{document}

\title{Electronic decoherence following photoionization: full quantum-dynamical treatment of the influence of nuclear motion}

\author{Caroline Arnold}
\email{caroline.arnold@cfel.de}
\affiliation{Center for Free-Electron Laser Science, DESY, Notkestrasse 85, 22607 Hamburg, Germany}
\affiliation{Department of Physics, University of Hamburg, Jungiusstrasse 9, 20355 Hamburg, Germany}
\affiliation{The Hamburg Centre for Ultrafast Imaging, Luruper Chaussee 149, 22761 Hamburg, Germany}

\author{Oriol Vendrell}
\email{oriol.vendrell@phys.au.dk}
\affiliation{Center for Free-Electron Laser Science, DESY, Notkestrasse 85, 22607 Hamburg, Germany}
\affiliation{The Hamburg Centre for Ultrafast Imaging, Luruper Chaussee 149, 22761 Hamburg, Germany}
\affiliation{Department of Physics and Astronomy, Aarhus University, Ny Munkegade 120, 8000 Aarhus, Denmark}

\author{Robin Santra}
\affiliation{Center for Free-Electron Laser Science, DESY, Notkestrasse 85, 22607 Hamburg, Germany}
\affiliation{Department of Physics, University of Hamburg, Jungiusstrasse 9, 20355 Hamburg, Germany}
\affiliation{The Hamburg Centre for Ultrafast Imaging, Luruper Chaussee 149, 22761 Hamburg, Germany}

\begin{abstract}
    Photoionization using attosecond pulses can lead to the formation of coherent superpositions of the electronic states of the parent ion. However, ultrafast electron ejection triggers not only electronic but also nuclear dynamics---leading to electronic decoherence, which is typically neglected on time scales up to tens of femtoseconds.  We propose a full quantum-dynamical treatment of nuclear motion in an adiabatic framework, where nuclear wavepackets move on adiabatic potential energy surfaces expanded up to second order at the Franck-Condon point.  We show that electronic decoherence is caused by the interplay of a large number of nuclear degrees of freedom and by the relative topology of the potential energy surfaces.  Application to $\chem{H_2O}$, paraxylene, and phenylalanine shows that an initially coherent state evolves to an electronically mixed state within just a few femtoseconds. In these examples the fast vibrations involving hydrogen atoms do not affect electronic coherence at short times. Conversely, vibrational modes involving the whole molecular skeleton, which are slow in the ground electronic state, quickly destroy it upon photoionization.
\end{abstract}

\maketitle


\section{\label{sect:introduction}Introduction}

The advent of attosecond pulses allows studying electronic correlation and ultrafast molecular dynamics through pump-probe experiments with unprecedented time resolution \cite{hen01, roz13, cal14}. Attosecond, broad-band pulses can be used to generate coherent superpositions of cationic states \cite{cal14, pab11, tim14, kra13}. The ionization triggers electronic and nuclear dynamics, finally leading to electronic decoherence. Decoherence is defined as the process where an initially pure state evolves to a statistical ensemble \cite{zhu04, vac14}. Electronic coherences are assumed to be responsible for a number of chemical processes, for example the high quantum efficiency of the energy conversion in photosynthesis, which is a matter of intensive current debate \cite{rom14}. 

Long-lived coherences are predicted by theories that focus on the evolution of the electronic subsystem, driven by electronic correlation \cite{kul07, gol15}. Nuclear motion is neglected, because the electrons move much faster than the heavier nuclei. This results in charge migration, an oscillatory motion of hole and electron density with frequencies defined by the energy gaps among the cationic states populated in the ionization process. In Ref.~\cite{cal14}, ultrafast dynamics in polyatomic molecules on a time scale shorter than the vibrational response were attributed to charge migration.

In recent theoretical work, the quantum nature of the nuclei was approximately taken into account \cite{vac15, vac15a, jen16, jen16a}. The authors sampled nuclear geometries within the width of the Gaussian wave packet of the nuclear ground state. This leads to a superposition of coherent oscillations with different frequencies that average out within a few femtoseconds. The cancellation of the  oscillations is due to the energy gap between the cationic states at the respective nuclear geometries. This approach takes into account the spatial delocalization of the nuclear wave packet. However, it does not consider the time evolution of the nuclear wave packet on the different potential energy surfaces. In the context of high-order harmonic generation (HHG) in molecules, the propagation of vibrational wave packets on adiabatic potential energy surfaces was studied in Ref.~\cite{pat09}. It was pointed out that the spatial overlap of vibrational wave packets in different electronic states depends on the topology of the potential energy surfaces. 

In this paper, we present a model for an \textit{ab-initio}, full quantum-mechanical treatment of nuclear motion on short time scales. Nuclei move on adiabatic potential energy surfaces that are expanded up to second order around the Franck-Condon point, i.e., the equilibrium geometry of the ground state. Bilinear mode-mode couplings are included. Due to the different values of the mode-mode couplings on each potential energy surface, they cannot be all removed simultaneously by rotation to another set of normal modes, and the model must be solved numerically. Using the multi-configuration time-dependent Hartree method (MCTDH) \cite{mctdh:MLpackage, bec00, mctdh1990} for the wave packet propagation, this allows us to numerically study the electronic coherence in large molecules considering all internal degrees of freedom at short times, as long as the Taylor expansion of the potential energy surfaces remains valid.
We observe electronic decoherence on a femtosecond time scale that can be attributed to the interplay of various vibrational modes in a molecule. Contrary to the widely held belief that fast modes are responsible for decoherence at short times \cite{lep14}, it is the small displacement of many slow modes that is ultimately responsible for the loss of coherence. In Sec.~\ref{sect:adiabatic-model} we introduce the model and argue that, generally, electronic decoherence is a multi-modal effect. This is illustrated in Sec.~\ref{sect:results} by application to $\chem{H_2O}$, paraxylene \cite{vac15a}, and phenylalanine \cite{cal14}.


\section{\label{sect:adiabatic-model}Model for the dynamics following photoionization}


We study the dynamics following photoionization in an electronically adiabatic framework. In the Born-Oppenheimer approximation, the nuclei move on adiabatic potential energy surfaces defined by the spectrum of the electronic Hamiltonian at the respective nuclear geometries. We consider the situation where a coherent attosecond pulse is used to excite a molecule above the ionization threshold. As the pulse is wide in the frequency domain, it is likely that two or more cationic states are populated in a coherent way \cite{pab11}. We assume vertical excitation, where the ground state nuclear wave packet is placed on the potential energy surfaces of the cationic electronic states. Furthermore, we neglect the photoelectron, so that the cation is initially in a pure electronic state.

Generally speaking, the equilibrium geometry of the ground state does not correspond to an equilibrium geometry of the excited states. After photoionization, the system enters a nonstationary state, and electronic and nuclear dynamics set in. In the adiabatic approximation, the nuclear wave packets move independently of each other on their respective potential energy surfaces. 


\subsection{\label{sect:hamiltonian}Adiabatic Hamiltonian}

Finding the shape of the adiabatic potential energy surfaces is the bottleneck towards establishing an exact solution of the problem outlined in the previous paragraph. The computational effort for determining the potential energy surfaces grows exponentially with the number of degrees of freedom. For small times, the potential energy surfaces can be approximated by Taylor polynomials up to second order, as proposed in the vibronic-coupling Hamiltonian \cite{wor04}. This corresponds to vibrations of the nuclei around their equilibrium configuration \cite{pat09}. In this work, we neglect nonadiabatic couplings, which will play a role only if the cationic electronic states feature conical intersections or avoided crossings close to the Franck-Condon region \cite{tim14}. The shorter and thus spectrally broader the ionization pulse, the less important nonadiabatic effects should become. The Hamiltonian of the cationic electronic state $\mu$ in a molecule with $f$ nuclear degrees of freedom is then given by

\begin{widetext}
\begin{equation}
    H^{(\mu)}(Q_1,\cdots,Q_f) = T + \Delta E^{(\mu)} 
    + \sum\limits_{i=1}^{f}\kappa_i^{(\mu)} Q_i 
    + \frac 1 2 \sum\limits_{i,j=1}^{f}\gamma_{ij}^{(\mu)} Q_i Q_j.
    \label{eq:hamiltonian-apes}
\end{equation}
\end{widetext}

We use atomic units and mass- and frequency-weighted normal-mode coordinates $Q_i$ throughout this paper. $T = \sum\limits_{i=1}^f - \frac{\omega_i}{2} \partial_{Q_i}^2$ refers to the kinetic energy, where the ground-state normal-mode frequencies $\omega_i$ are employed. The energy difference to the lowest cationic state is denoted by $\Delta E^{(\mu)}$. The coefficients $\kappa_i^{(\mu)}, \gamma_{ii}^{(\mu)}, \gamma_{ij}^{(\mu)}$ are obtained by central-difference approximation for small displacements from the Franck-Condon point along the normal modes. They describe the gradient, curvature, and mode-mode coupling, respectively. 

For propagating the nuclear wave packet, we use the multi-configuration time-dependent Hartree method (MCTDH) \cite{mctdh1990} in its multi-set formulation \cite{fan94, wor96}. The nuclear wave packet in each electronic state is described by an independent product of time-dependent single particle functions, thus optimally treating the independent nuclear evolution on the different cationic potential energy surfaces.



\subsection{\label{sect:electronic-decoherence}Electronic Decoherence}

\begin{figure}[h]
  \centering
  \includegraphics[width=\columnwidth]{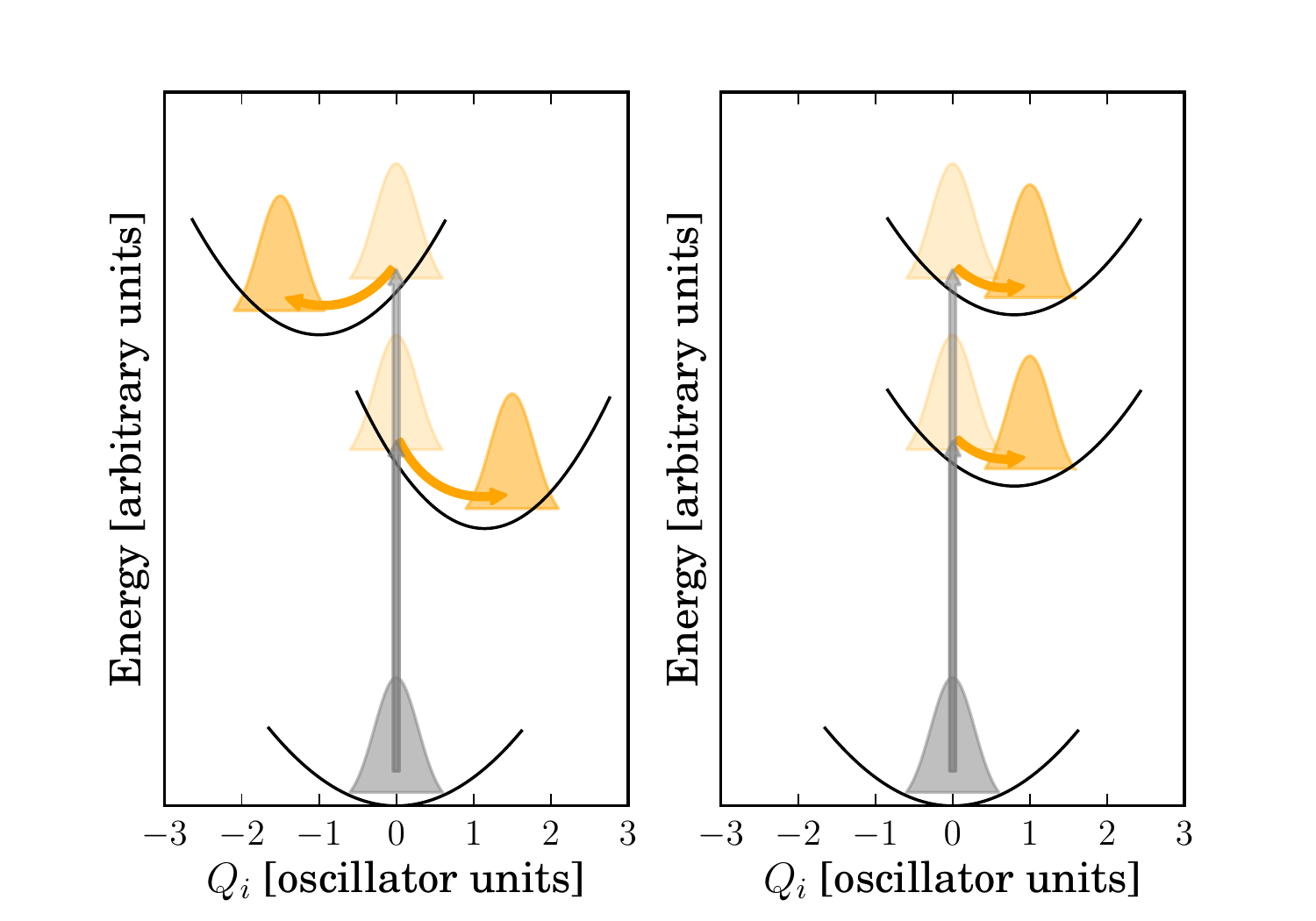}
  \caption{The spatial overlap of vibrational wave packets, and thus the electronic coherence, depends on the topology of the potential energy surfaces relative to each other. This is illustrated here for the motion along one normal-mode coordinate $Q_i$. After excitation from the ground state (gray), the nuclear wave packets evolve on the cationic potential energy surface (orange). Left side: Different gradients and curvatures lead to diminishing spatial overlap and decoherence. Right side: The potential energy surfaces are displaced only vertically. The nuclear wave packets move synchronously; spatial overlap is high throughout the propagation.}
  \label{fig:topology-sketch}
\end{figure}

To quantify electronic decoherence, the coherence among the $n$ cationic states is calculated from the reduced density matrix $\rho$ of the electronic subsystem. Starting from the Born-Huang ansatz \cite{bor54}, the decomposition of the total wave function in terms of nuclear and adiabatic electronic wavefunctions $\chi, \phi$, respectively, reads

\begin{equation}
  \Psi(r,Q,t) = \sum\limits_{\mu = 1}^n c_\mu \chi_\mu(Q, t) \phi_\mu(r; Q),
  \label{eq:born-huang}
\end{equation}
where $r$ refers to electronic coordinates. The expectation value of an electronic observable $\hat O (r)$ can be obtained via

\begin{align}
  \langle \hat O (r) \rangle
  &
  =
  \int \mathrm{d} Q \int \mathrm{d} r \Psi^*(r, Q, t) \hat O (r) \Psi(r, Q, t) \\
  &
  = 
  \sum\limits_{\mu\nu} \int \mathrm{d} Q c_\mu^{*} \chi_\mu^{*} (Q, t) O_{\mu\nu} (Q)c_\nu \chi_\nu (Q, t) \\
  &
  \approx
  \sum\limits_{\mu\nu} O_{\mu\nu} \rho_{\nu\mu},
  \label{eq:electronic-observable-expectation-value}
\end{align}
where for the last line we make the assumption $O_{\mu\nu}(Q) = O_{\mu\nu}$. The electronic reduced density matrix elements in Eq.~\eqref{eq:electronic-observable-expectation-value} correspond to the nuclear wave packet overlaps:

\begin{equation}
  \rho_{\nu\mu} (t)
  = 
  c_\nu c_\mu^{*} \int \mathrm{d} Q \chi_0^*(Q, 0) \mathrm{e}^{i H^{(\mu)} t} \mathrm{e}^{- i H^{(\nu)} t} \chi_0(Q, 0).
  \label{eq:nuclear-wave-packet-overlap}
\end{equation}

In the single-geometry model \cite{vac15a, jen16a}, the nuclear functions reduce to delta peaks, $\chi_0(Q) = c_0\delta(Q - Q_0)$. Neglecting nuclear motion, Eq.~\eqref{eq:nuclear-wave-packet-overlap} reduces to the long-lived electronic coherences predicted in charge-migration theories \cite{cal14, raa99, bre03, gol15}

\begin{equation}
  \tilde \rho_{\nu\mu}
  =
  c_\nu^* c_\mu \mathrm{e}^{i\left(\Delta E^{(\mu)} - \Delta E^{(\nu)}\right)t}.
  \label{eq:t-0}
\end{equation}
Evolution of this single geometry along a trajectory as a consequence of an averaged force corresponds to Ehrenfest dynamics, also resulting in a perpetually pure electronic subsystem without decoherence. A full quantum-mechanical treatment of the nuclei, however, reveals that electronic decoherence is caused by the fast spread of nuclear wave packets along all degrees of freedom. The associated speed is determined by the relative topology of the potential energy surfaces, as shown in Fig.~\ref{fig:topology-sketch}. Consider an initial state formed by a product of Gaussians and the electronic ground state $\ket{0}$ 

\begin{equation}
  \Psi_0
  =
  \prod\limits_{i=1}^{f} \chi_{0, i} \ket{0},
  \quad
  \chi_{0, i}
  = 
  \mathcal{N}_i \mathrm{e}^{- \left(Q_i - Q_{0,i}\right)^2 / 2}
  \label{eq:ground-state-product-gaussians}
\end{equation}
with appropriate prefactors $\mathcal{N}_i$ ensuring normalization. This ground-state wave packet is placed on the cationic surfaces following the ionization by a coherent pulse. Neglecting the mode-mode couplings $\gamma_{ij}^{ (\mu)}, i \neq j$, the Hamiltonian allows one to factorize the wave packet along the coordinates $Q_i$ for all times, where $H^{(\mu)} = \sum_{i=1}^f H_i^{(\mu)}$. From Eq.~\eqref{eq:nuclear-wave-packet-overlap}, 

\begin{widetext}
\begin{equation}
  \rho_{\nu\mu} (t)
  =
  c_\nu c_\mu^{*} \prod\limits_{i=1}^{f} 
  \int \mathrm{d}Q_i \chi_{0, i}^{*} (Q_i, 0) \mathrm{e}^{i H_{i}^{(\mu)} t} \mathrm{e}^{-i H_{i}^{(\nu)} t} \chi_{0,i} (Q_i, 0)
  =
  c_\nu c_\mu^{*} \prod\limits_{i=1}^f \rho_{\nu\mu}^{(i)}(t).
  \label{eq:coherence-product-oscillating-factors}
\end{equation}
\end{widetext}
The oscillating $\rho_{\nu\mu}^{(i)}(t)$ will dephase, because, in general, the frequencies defined by the gaps among the cationic electronic states are $Q$-dependent. Essentially, coherence is lost because the product of a large number of factors ranging between 0 and 1 tends to 0. The process is speeded up if at least one mode contributes a factor close to 0. This corresponds to a vanishing spatial overlap caused by differing gradients in the potential energy surfaces. The frequency of the mode itself in the ground state does not determine its influence on decoherence. 

%

We calculate the purity $\mathrm{Tr}(\rho^2)$ to monitor the evolution of an initially pure state into a mixed state. The former yields $\mathrm{Tr}(\rho^2)=1$, the latter, in an $n$-state system with equal weights $\frac 1 n$ for each state, $\mathrm{Tr}(\rho^2)=1/n$.  This approach is convenient especially if more than two cationic states are considered; furthermore, it is representation independent. Note that if the decay of the off-diagonal matrix element can be expressed as $\rho_{\mu\nu} = c \mathrm{e}^{- \gamma t}$, the purity decays twice as fast; for an equally weighted two-level system, $\mathrm{Tr}(\rho^2) = \frac 1 2 + 2 |c|^2 \mathrm{e}^{- 2 \gamma t}$.


\section{\label{sect:results}Results and discussion}

In the following, we apply our model to $\chem{H_2O}$, paraxylene, and phenylalanine cations.  For $\chem{H_2O}$ and paraxylene, the potential energy surfaces were calculated at the multi-configurational self-consistent field (MCSCF) level of theory restricting the active orbitals of the ion to occupied orbitals in a Hartree-Fock calculation of the neutral system and using the GAMESS software package \cite{gamess1993}. This level of accuracy was found to be sufficient in previous work \cite{li13, li14}.  For phenylalanine, we approximate the potential energy surfaces of the cation using a single configuration, using Koopmans' theorem. We compare one-dimensional simulations, where only one normal mode at a time is considered, to full-dimensional simulations. By this, we demonstrate that decoherence is an effect that can be attributed to the interplay of a large number of vibrational modes, rather than to a small number of fast modes.

\subsection{\label{sect:results:h2o}\chem{H_2O}}

\begin{figure}[h]
    \centering
    \includegraphics[width=\columnwidth]{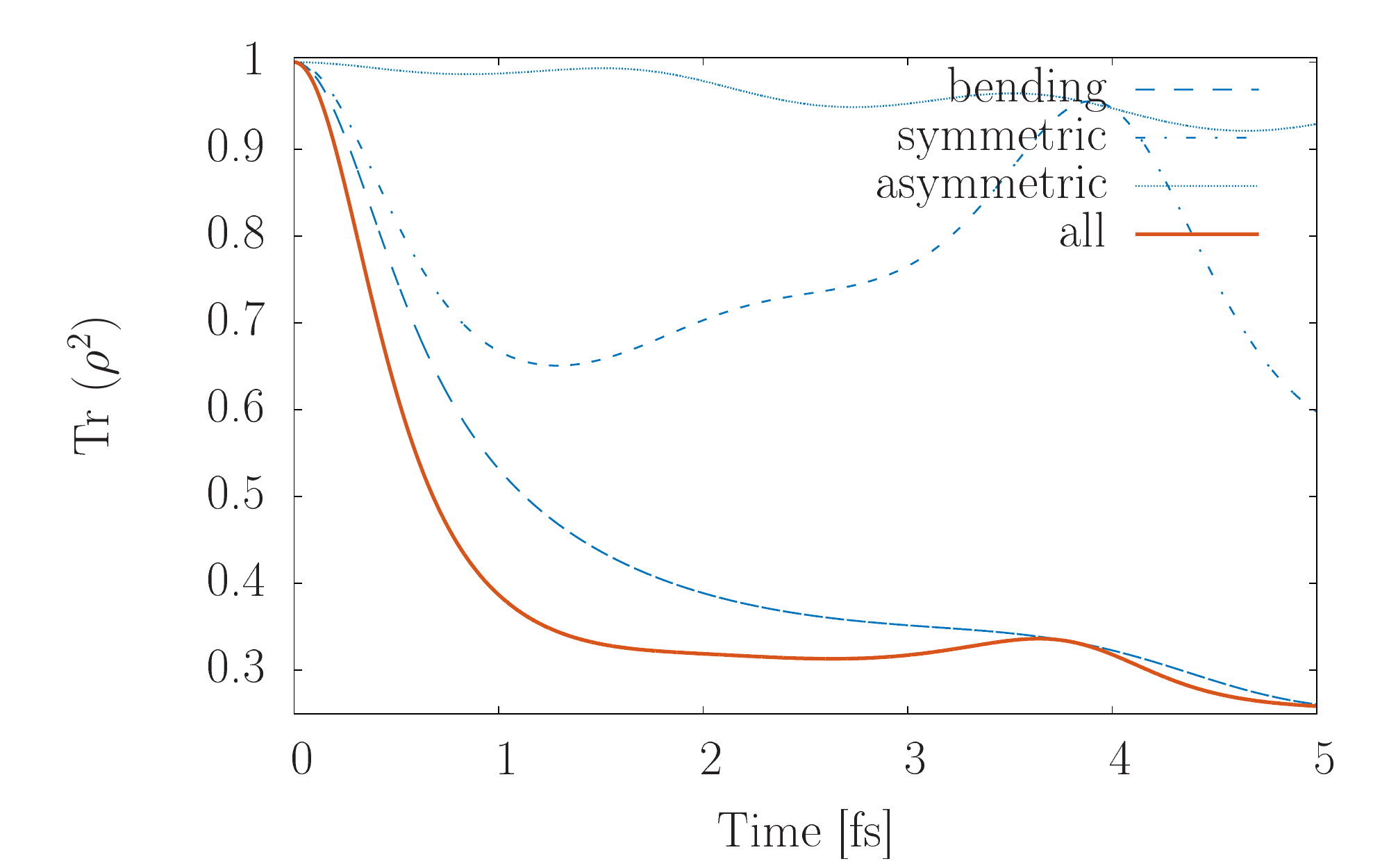}
    \caption{Evolution of electronic purity in $\chem{H_2O}$ with four cationic states. The propagation using one mode only (blue dashed lines) is compared to the propagation of the full-dimensional system (orange solid line). The system, initially prepared in an equally weighted superposition of all cationic states, evolves to a mixed state within $\unit[1]{fs}$. This is attributed to the loss of coherence along the bending mode as well as the interplay of the three vibrational modes.  }
    \label{fig:h2o-single-combined}
\end{figure}

$\chem{H_2 O}$ was chosen as a model molecule for its simplicity. The coefficients for the adiabatic potential energy surfaces were calculated for four cationic states using seven electrons in four active orbitals.  The energy spacings with respect to the cation ground state are \unit[0.06]{eV}, \unit[0.27]{eV}, \unit[0.87]{eV}, respectively.

Figure \ref{fig:h2o-single-combined} shows the evolution of the coherence in the $\chem{H_2O}$ molecule. The results for one-dimensional simulations with one normal mode each are shown together with the simulation of the full three-dimensional system. 

We observe that the initially pure state 

\begin{widetext}
\begin{equation}
    \Psi(\vect Q, 0) 
    =
    \chi_{0,1}(Q_1, 0) \chi_{0,2}(Q_2, 0) \chi_{0,3}(Q_3, 0)
    \frac{1}{\sqrt 4} \left(\ket{1} + \ket{2} + \ket{3} + \ket{4}\right)
    \label{eq:h2o-initial-state}
\end{equation}
\end{widetext}
evolves towards a mixture in the electronic subsystem. A constant value is reached already after \unit[1]{fs}. The individual modes are identified with the asymmetric stretch ($\unit[3756]{cm^{-1}}$), the symmetric stretch ($\unit[3657]{ cm^{-1} }$) and the bending motion ($\unit[1595]{cm^{-1}}$). While the fast modes alone maintain electronic coherence, the slow mode is mainly responsible for the overall loss of coherence. The rate of the decoherence, and the suppression of recurrence, is attributed to the dephasing of the oscillations along single modes. Even in $\chem{H_2 O}$ with only three degrees of freedom, the loss of coherence is thus seen to be a multi-modal effect with a strong participation of the slower bending motion.

The full quantum-mechanical treatment of nuclear motion reveals its high influence on electronic decoherence. In Ref.~\cite{li13a}, long-lived oscillations were observed in ionized liquid water and were attributed to coherent hole dynamics including the two lowest-lying cationic states. While the chemical environment of a single water molecule does not directly translate to liquid water or water clusters, the loss of coherence can be expected to occur even faster if more modes are included. Restriction to two states in $\chem{H_2O}$ does not change the decoherence time scale.

\subsection{\label{sect:results:paraxylene}Paraxylene}

\begin{figure}[h]
    \centering
    \includegraphics[width=\columnwidth]{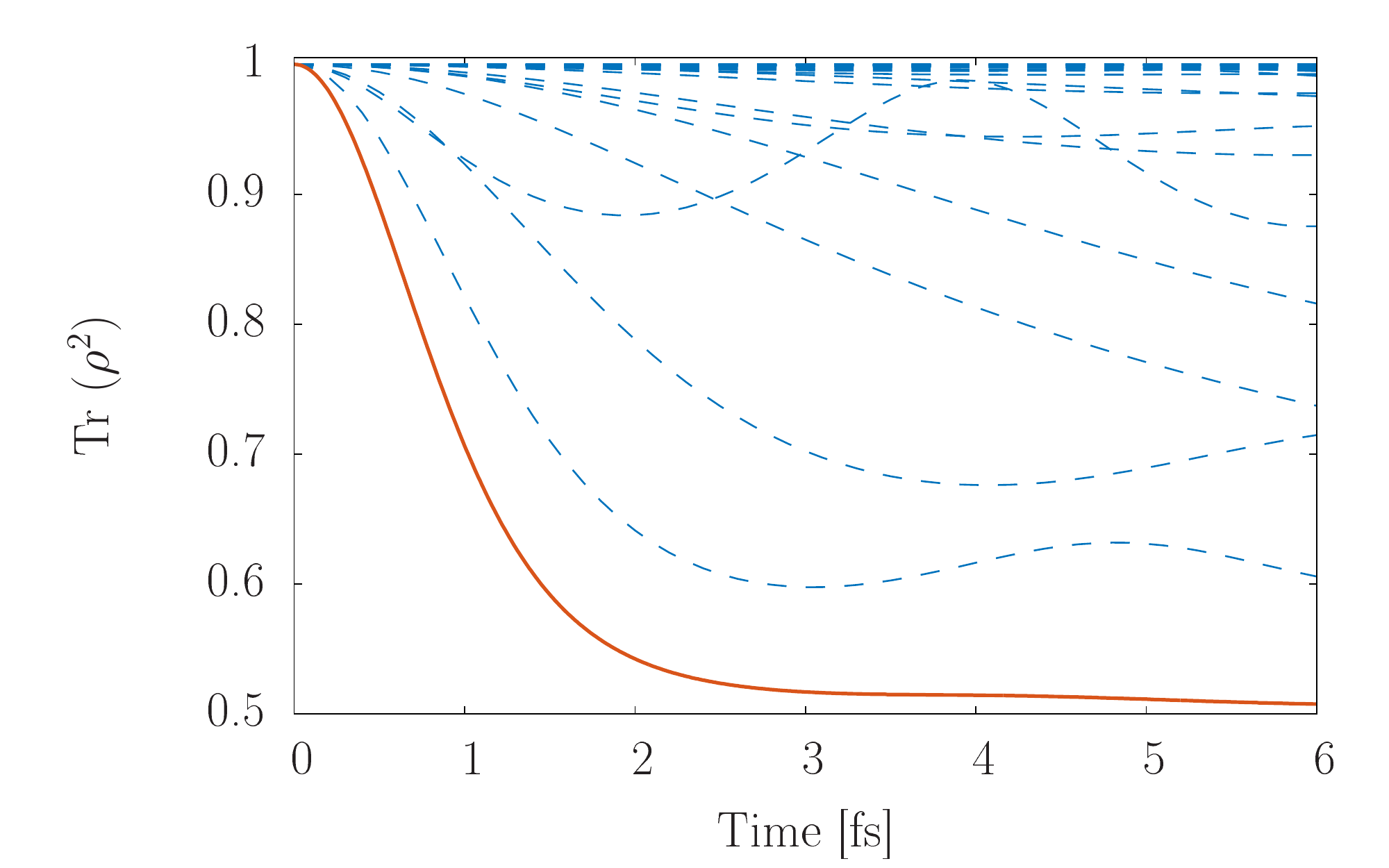}
    \caption{%
      Evolution of electronic purity in paraxylene with two cationic states. One-dimensional simulations (blue dotted lines) are compared to the propagation of the full-dimensional system (orange solid line). The interplay of the large number of modes leads to a mixed state on a time scale of $\unit[2-3]{fs}$. Recurrence can be seen in the one-dimensional simulations, but is suppressed when all vibrational modes are taken into account. 
    }
    \label{fig:paraxylene-single-combined}
\end{figure}

\begin{figure}[h]
    \centering
    \includegraphics[width=\columnwidth]{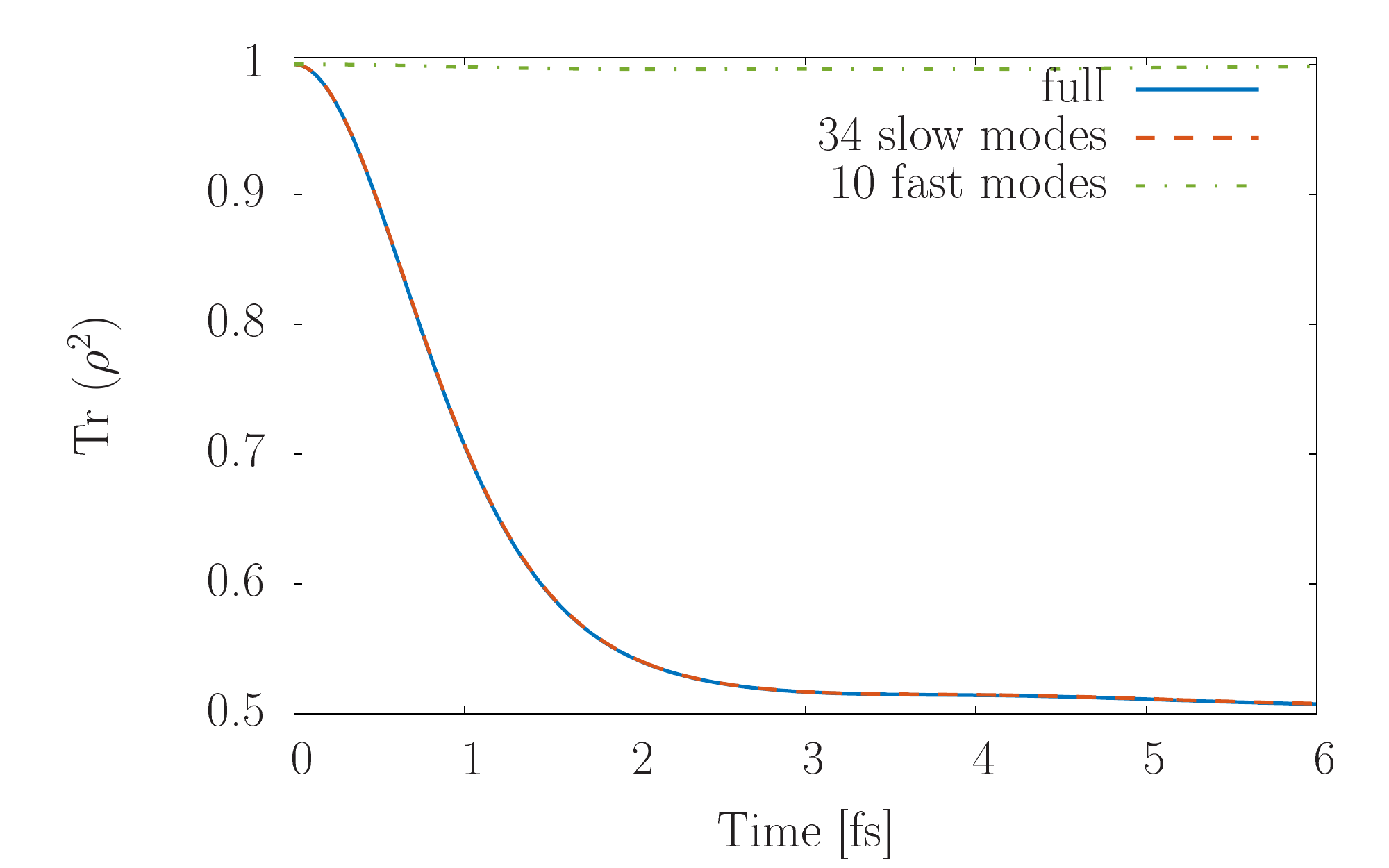}
    \caption{Electronic coherence is reduced by the interplay of all vibrational modes. The ten fastest modes, corresponding to the C-H vibrations, maintain coherence. It is the slow modes, corresponding to vibrations of the whole molecule and the benzene-ring carbon atoms, that account for the fast decoherence. }
    \label{fig:paraxylene-speed}
\end{figure}

\begin{figure}
    \centering
    \includegraphics[width=\columnwidth]{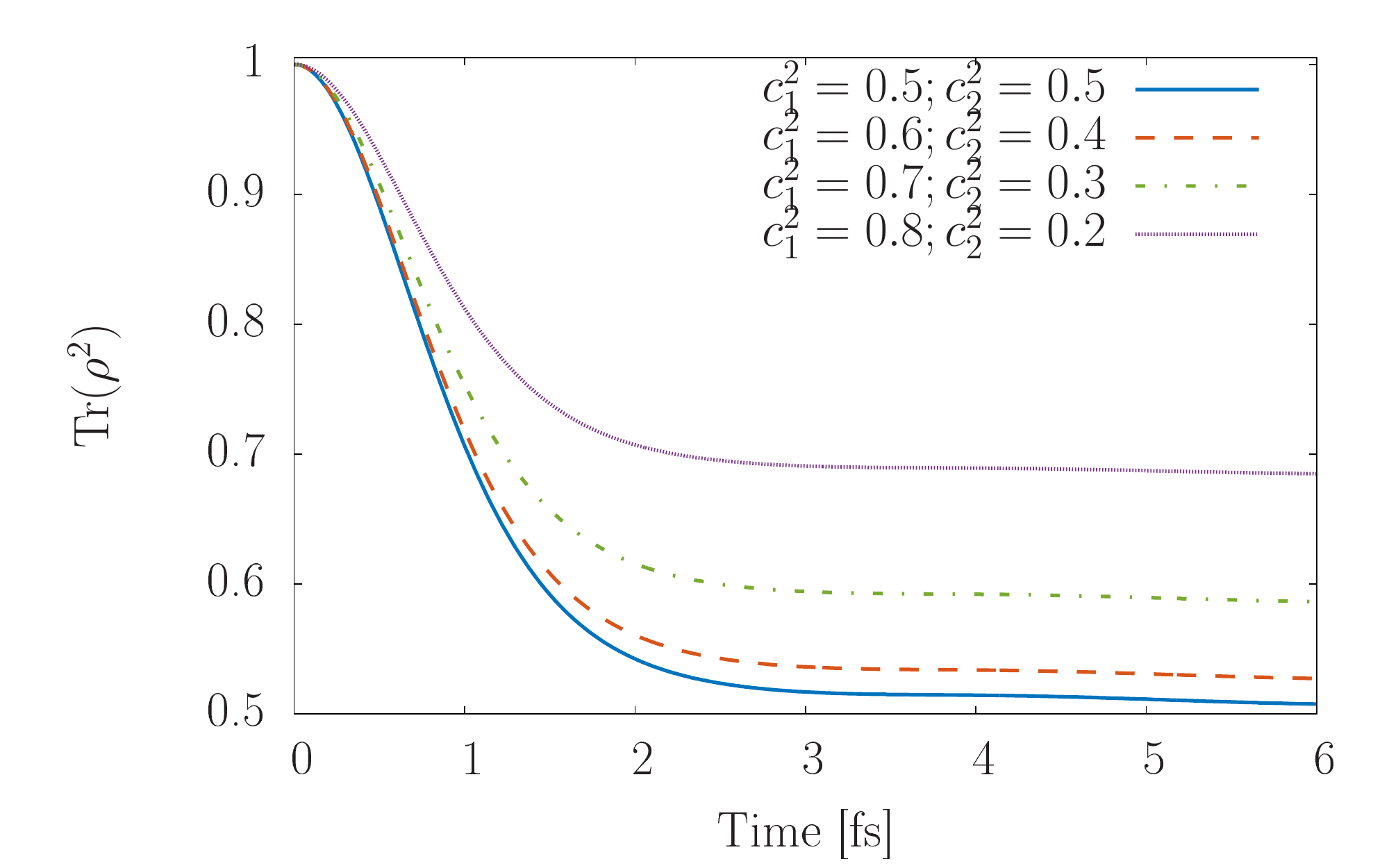}
    \caption{%
        Electronic decoherence in paraxylene for an initially pure state $\Psi(\vect Q, 0) = \chi_0(\vect Q, 0) \left(\sqrt{c_1} \ket{1} + \sqrt{c_2} \ket{2}\right)$. The time scale where electronic decoherence is destroyed is independent of the choice of weights $c_1, c_2$. A mixed state is reached within few femtoseconds, where $\mathrm{Tr}\left(\rho^2\right) = c_1^4 + c_2^4$.
    }
    \label{fig:paraxylene-weights}
\end{figure}

\begin{figure}[h]
  \centering
  \includegraphics[width=\columnwidth]{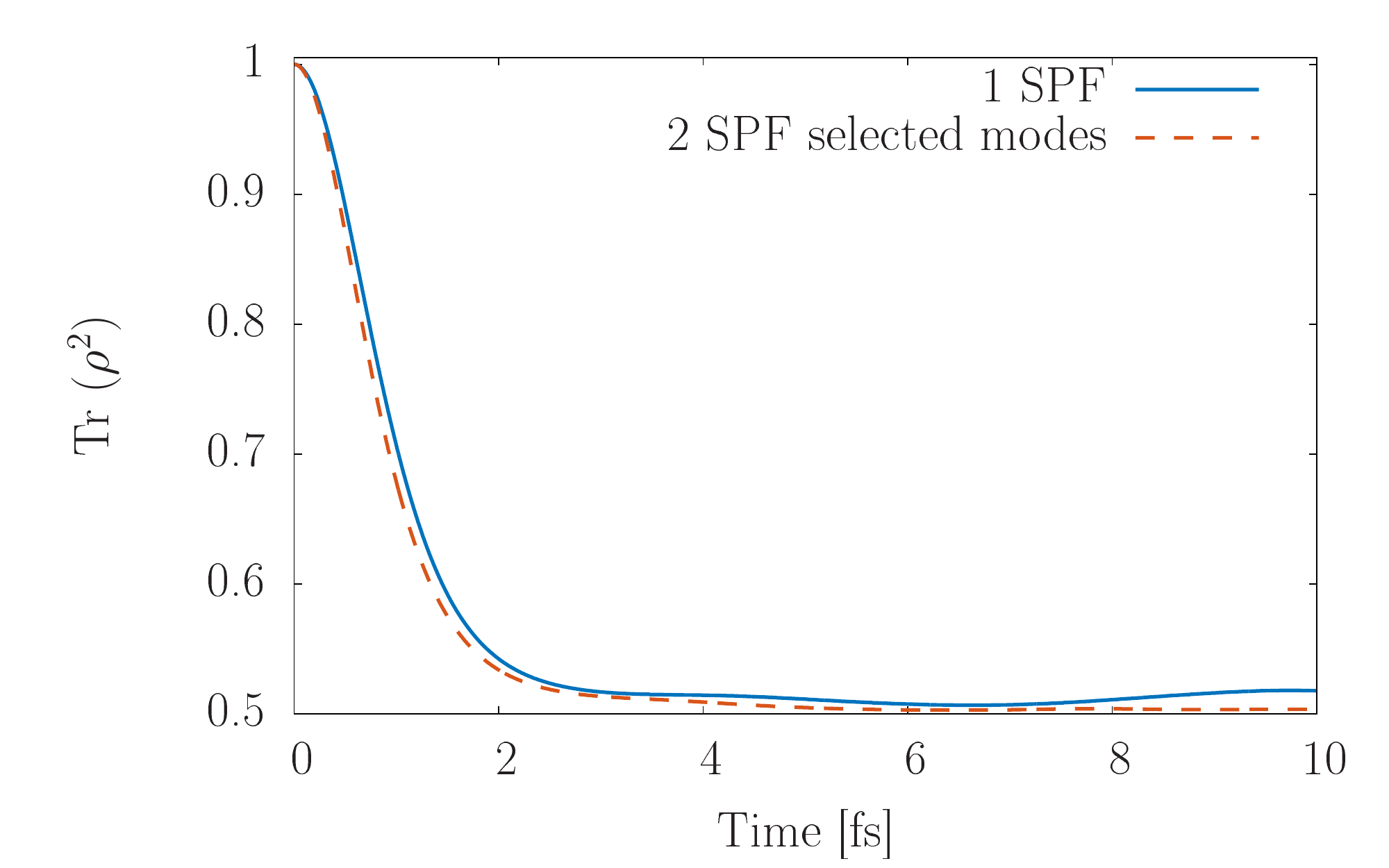}
  \caption{%
  In paraxylene, a subset of 12 modes that are coupled to each other $(|\gamma_{ij}^{(\mu)}| > 10^{-3})$ is given a second single particle function per mode (orange dotted line), compared to one SPF per mode (blue solid line). This suppresses the small recurrence at $\unit[8-10]{fs}$.}
  \label{fig:paraxylene-spf}
\end{figure}

\begin{figure}
  \centering
  \includegraphics[width=\columnwidth]{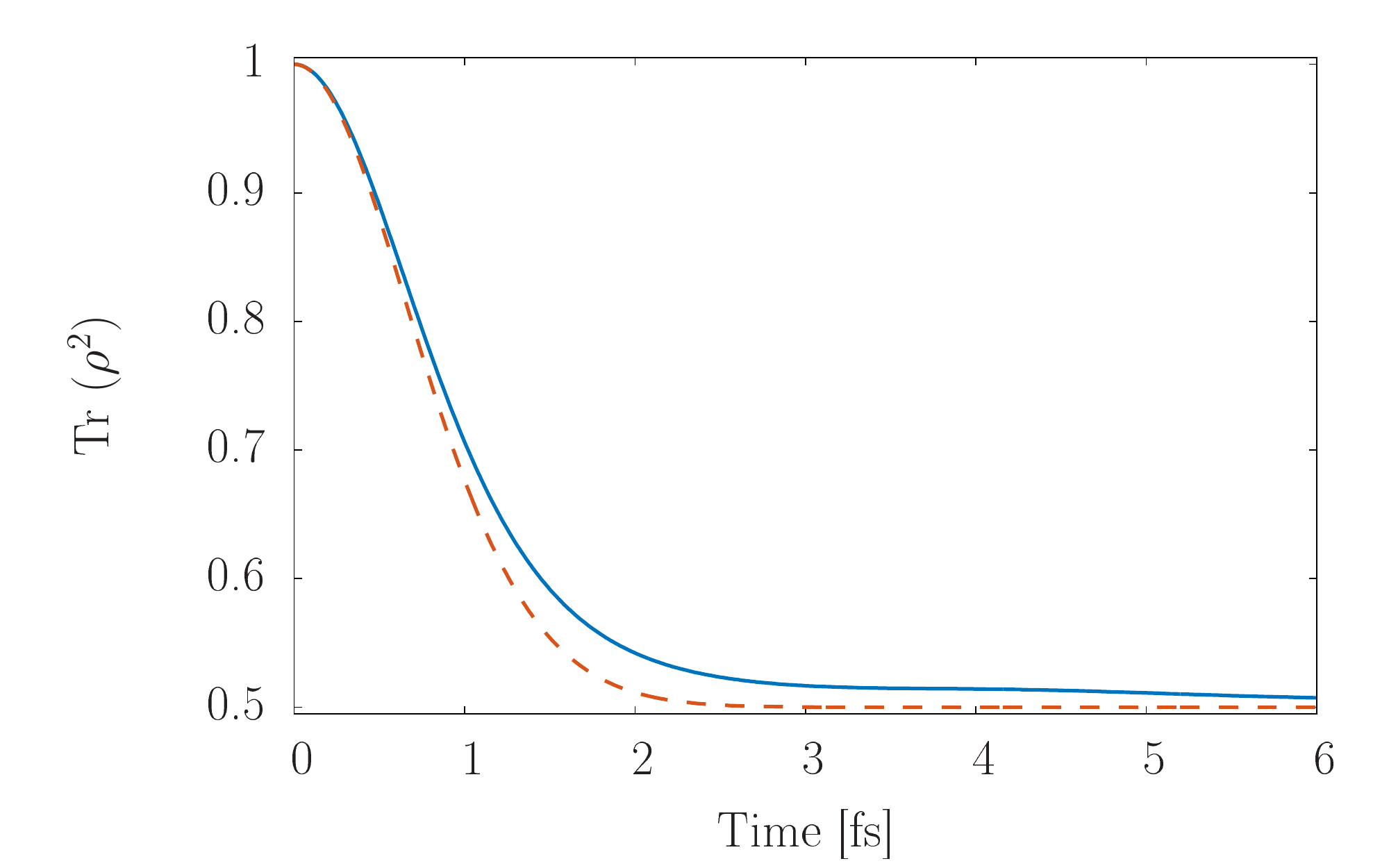}
  \caption{%
  Electronic decoherence in paraxylene including (blue solid line) and neglecting (orange dashed line) the kinetic energy operator in the quantum-dynamical calculation.
  }
  \label{fig:paraxylene-kinetic}
\end{figure}

We now apply the model to paraxylene, a benzene molecule with two methyl groups on opposite sides of the benzene ring. Paraxylene was included in the aforementioned study \cite{vac15a}, which makes it suitable to compare the conceptually different approaches. With 18 atoms, paraxylene has 48 internal degrees of freedom. We chose to project out the two methyl group rotations, which otherwise should have been treated as hindered rotors with a periodicity of $60^\circ$ \cite{ven02}. Two additional very low frequency modes were left out of the model to avoid numerical problems. Leaving modes out of the model can only slow down decoherence, not accelerate it, thus leaving our conclusions unchanged. 

Following Ref.~\cite{vac15a}, we include the first two cationic states of paraxylene, separated by $\unit[0.51]{eV}$ at the Franck-Condon point. The cationic potential energy surfaces were determined at the MCSCF level of theory using five electrons in three active orbitals.

Initially, the system is prepared in an equally weighted superposition of the two cationic states,

\begin{equation}
  \Psi(\vect Q, 0)
  =
  \prod\limits_{i=1}^{48} \chi_{0,i}(Q_i, 0) \frac{1}{\sqrt 2} \left(\ket{1} + \ket{2}\right).
  \label{eq:paraxylene-coherent-ground-state}
\end{equation}
The high number of degrees of freedom restricts us to one time-dependent single particle function (SPF) per degree of freedom and state, which is however fully flexible and expanded in a harmonic DVR grid \cite{bec00}. For selected states, more SPF were included to estimate their contributions. 

Figure \ref{fig:paraxylene-single-combined} shows the evolution of the initially pure state to a mixed state on the timescale of $\unit[2-3]{fs}$. The one-dimensional simulations, represented by the blue dashed curves, show that coherence is maintained at the single-mode level for most of the modes. Recurrence is observed after the associated oscillation period. The overall coherence, as outlined in Sec.~\ref{sect:adiabatic-model}, can be described as the product of the one-dimensional coherences, if mode-mode couplings are neglected. Due to the dephasing of the oscillations along the different normal modes, the overall coherence drops to zero and does not recur. This shows that the loss of coherence can be explained from the high number of degrees of freedom. There is no single mode that is responsible alone for the fast decoherence. The breathing mode $(\unit[1330]{cm^{-1}})$ and the two Kekule modes $(\unit[1777]{cm^{-1}}, \unit[1831]{cm^{-1}})$ show recurrence of the purity on a femtosecond timescale, if treated individually. 


In Fig.~\ref{fig:paraxylene-speed}, we study the influence of various selected modes on the decoherence. The ten fastest normal modes correspond to C-H vibrations. Electronic coherence is maintained if only this subset of modes is considered. The 34 slowest modes, with their relatively long recurrence times, lead instead to rapid loss of coherence. It is thus not the speed of the nuclear motion that defines the influence on electronic decoherence, but rather the topology of the potential energy surfaces relative to each other, as illustrated in  Fig.~\ref{fig:topology-sketch}. Note that, within our adiabatic model, the decoherence rate is independent of the weights assigned to the initial states, as shown in Fig.~\ref{fig:paraxylene-weights}.

In a typical MCTDH calculation, the results are taken to be converged if the population of the highest SPF is below $10^{-3}$, or if the effect of leaving the highest SPF out is below $10^{-5}$ (see Ref.~\cite{bec00}). This level of accuracy is certainly not feasible with a large number of degrees of freedom. However, we can approximate by the following reasoning: There are many pairs of vibrational modes without discernible mode-mode coupling. Along these modes, the nuclear wave packet is very well described using only one SPF. More SPFs are needed to capture the dynamics with respect to coupled modes. The one-dimensional calculations are of course exact with one SPF per surface.

Figure \ref{fig:paraxylene-spf} shows the effect of including a second SPF for a subset of twelve coupled modes $(|\gamma_{ij}^{(\mu)}| > 10^{-3})$. The results are comparable to the ones obtained by using only one SPF per mode. It is noticeable that the recurrence at $\unit[8]{fs}$ is suppressed if two SPF are used in this group of modes and that decoherence becomes slightly faster. 

In Ref.~\cite{vac15a}, electron dynamics following photoionization was studied in paraxylene. Decoherence was attributed to the width of the nuclear wave packet, because the energy gap and hence the frequency of $\rho_{12}$ changes over its extension. After averaging over all nuclear coordinates, coherence is lost without recurrence, in the case of paraxylene with a half-life time of $t_{1/2} = \unit[4]{fs}$. This approach does not allow for transfer of density between the different nuclear geometries within the wave packet and decoherence becomes a consequence of dephasing by averaging the coherent matrix elements over many different geometries. It follows from Eq.~\eqref{eq:nuclear-wave-packet-overlap} if the kinetic energy operator is neglected and the nuclear wavepacket is replaced by a geometry distribution $p(Q)$. Note that non-adiabatic effects are thereby excluded. Although this model provides a physically reasonable time scale for decoherence in multidimensional systems, it is unclear when and how the approximation breaks down and how to systematically improve it. In contrast, in the model proposed here, the whole nuclear wave packet evolves fully quantum-mechanically and the treatment of decoherence, and potentially of time-resolved electronic spetroscopies, becomes exact at short times. By comparing, within our model, the decoherence rate with and without including kinetic energy, as shown in Fig.~\ref{fig:paraxylene-kinetic}, we find that neglecting the kinetic energy operator in the quantum-dynamical calculation slightly overestimates electronic decoherence and misses the persisting linear component seen in the long-term behavior.

\subsection{\label{sect:results:phenylalanine}Phenylalanine}

\begin{figure}
    \centering
    \includegraphics[width=\columnwidth]{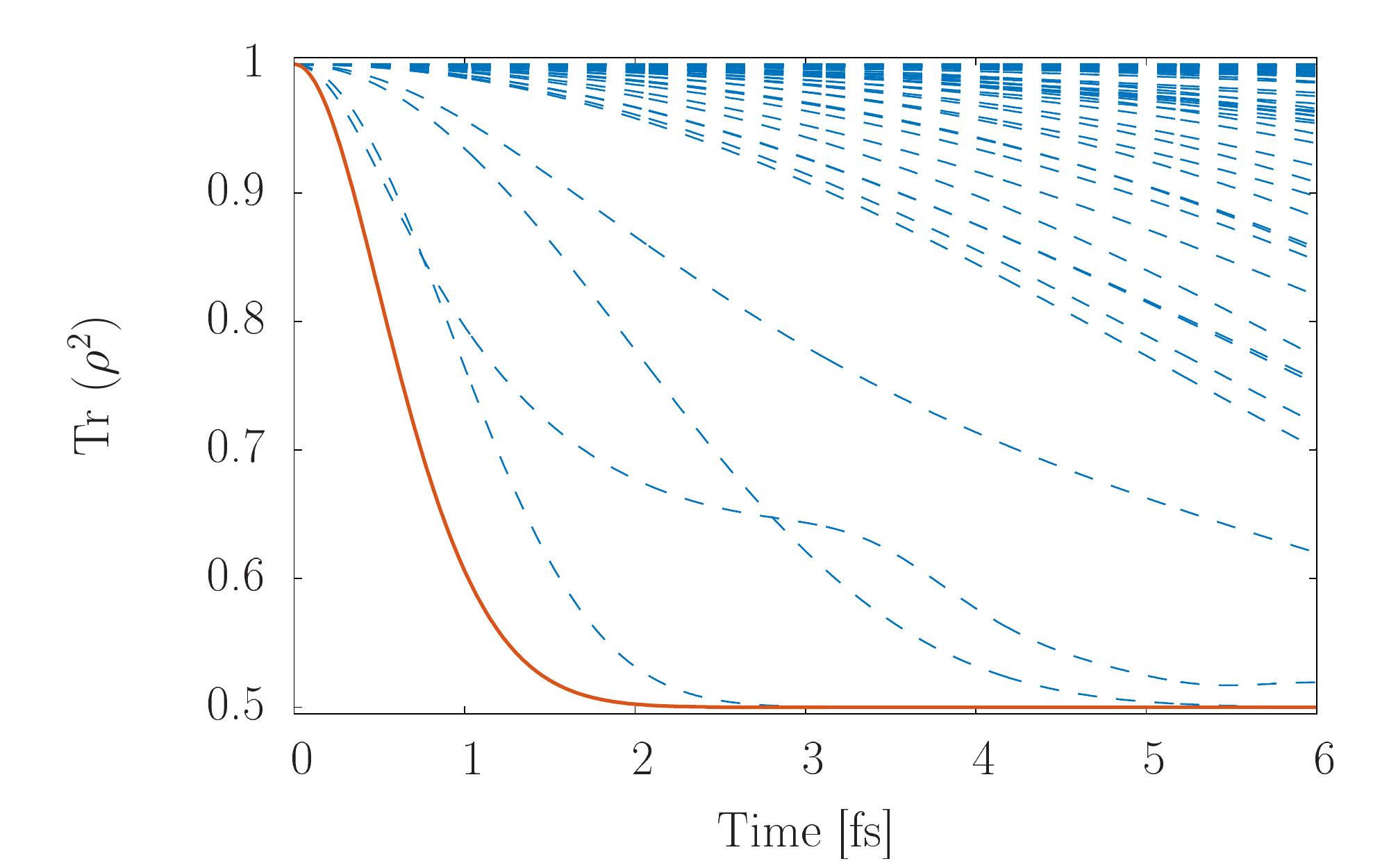}
    \caption{Evolution of electronic purity in phenylalanine with two cationic states. One-dimensional simulations for each of the 63 normal modes (blue dotted lines) are compared to the propagation including all modes (orange solid line) for an initially pure state. Electronic decoherence occurs within $\unit[1]{fs}$.}
    \label{fig:phenylalanine-trace-squared}
\end{figure}

In Fig.~\ref{fig:phenylalanine-trace-squared}, we show the evolution of electronic coherence in phenylalanine, which was studied in Ref.~\cite{cal14} for an initial state arising from the photoionization cross-sections. In our analysis, initially, the cation ground state and the first excited state are in a coherent superposition.  Again, we use one SPF per surface and mode, and we compare one-dimensional simulations to the propagation including all 63 degrees of freedom. Most mode-mode couplings have a value $|\gamma_{ij}^{(\mu)}| < 10^{-3}$. We find that the initially pure state evolves to a mixed state within about $\unit[1]{fs}$, if all modes are considered. Recurrence is suppressed, as the oscillations dephase along the different modes. While most modes preserve electronic coherence over the period of investigation, we observe a stronger participation of relatively slow vibrational modes with ground-state frequencies of $ \unit[1816]{cm^{-1}} , \unit[1809]{cm^{-1}}$ , and  $\unit[1340]{cm^{-1}} $, respectively.


\section{\label{sect:conclusion}Conclusion and Outlook}

It is often assumed that, if nuclear motion plays a role in electronic decoherence, it is the fast modes that should be considered \cite{cal14}. In an organic molecule, these are the C-H vibrations with a typical frequency of about $\unit[3000]{cm^{-1}}$, corresponding to a period of vibration of $\unit[11]{fs}$. This is typically slower than calculated charge migration times and is therefore not considered to have an effect. Our study, however, indicates that it is not the fast, but the interplay of the slow modes (in the sense of their inverse vibrational frequencies in the neutral ground state) that causes decoherence on a femtosecond timescale. In Fig.~\ref{fig:paraxylene-speed}, we show that, in paraxylene, if only the 10 C-H vibrations are considered, the system stays coherent within at least $\unit[6]{fs}$. This is because the potential energy surfaces for the fast modes turn out to be only vertically displaced from each other. As shown schematically in Fig.~\ref{fig:topology-sketch} and discussed in Sec.~\ref{sect:electronic-decoherence}, in this case, the nuclear wave packets move synchronously on the potential energy surfaces, and their spatial overlap does not change much. In the slower modes, the potential energy surfaces can also be horizontally displaced from each other, and the spatial overlap is reduced. 

We presented an \textit{ab-initio} model for electronic decoherence following photoionization that takes the quantum nature of the nuclei into account and allows for a full-dimensional treatment of the molecule. Combined with a proper treatment of the preparation step and with the consideration of the relevant electronic observables it can be used to interpret and predict the outcome of current experiments in molecular attoscience \cite{cal14}.  The probe step can be excluded from the analysis, as the loss of electronic coherence would be apparent in any probe technique sensitive to electronic structure.  We showed that electronic decoherence can be explained by considering the topologies of different potential energy surfaces. It is the interplay of a large number of vibrational modes that are causing decoherence in the electronic density matrix, not a set of e.g. fast C-H vibrations. Our results suggest that in molecular systems, purely electronic dynamics that may be described in terms of a coherent electronic wave packet, exists only for sub-femtosecond time scales, and nuclear motion cannot be neglected. With the approach shown in this paper, one can calculate the time-dependent density matrix of the electronic subsystem and from it the time-dependent expectation values of any observable that depends only on the electronic subsystem via
$
\langle A \rangle = \mathrm{Tr}(A \rho),
$
for example the hole density. Observables of the form $O_{\mu\nu}(Q)$ that depend on the electronic states and the nuclear geometry can be easily computed as well since the full nuclear wave packets are available. 

The model is currently limited to short time scales. In the future, we plan to explore quantum-classical approaches that allow us to propagate the nuclei for longer times. However, we note that for describing the short-time electronic response with nuclear-induced decoherence, this model (with possible non-adiabatic extensions) is all what is required, as decoherence sets in before the nuclear wave packets abandon the quadratically expanded region of the potential energy surfaces. The exact quantum-mechanical calculations will serve as a reference for future development of these methods. 


\begin{acknowledgments}
This work has been supported by the excellence cluster \textit{The Hamburg Centre for Ultrafast Imaging - Structure, Dynamics and Control of Matter at the Atomic Scale} of the Deutsche Forschungsgemeinschaft.
\end{acknowledgments}



%


\end{document}